\documentclass[a4paper,11pt]{article}
\usepackage{pos}
\usepackage{amsmath,leftidx}
\usepackage{dsfont}
\usepackage{bbm}
\usepackage{cleveref}
\usepackage{subcaption}
\usepackage[utf8]{inputenc}
\usepackage{mathrsfs}
\usepackage{amsfonts}
\usepackage{wrapfig}
\usepackage{enumitem}

\let\OLDthebibliography\thebibliography
\renewcommand\thebibliography[1]{
  \OLDthebibliography{#1}
  \setlength{\parskip}{0pt}
  \setlength{\itemsep}{0pt plus 0.3ex}
}

\newcommand{\fm}{\text{fm}}

\title{A strategy for B-physics observables in the continuum limit}
\ShortTitle{A strategy for B-physics observables in the continuum limit}

\author[ma,mz,g]{Alessandro Conigli}
\author[n]{Julien Frison}
\author[d]{Patrick Fritzsch}
\author[mar]{Antoine G\'erardin}
\author[mu]{Jochen Heitger}
\author[ma]{Gregorio Herdoiza}
\author[c,mz,g]{Simon Kuberski}
\author[ma]{Carlos Pena}
\author[n]{Hubert Simma}
\author*[n,h]{Rainer Sommer}

\affiliation[n]{John von Neumann-Institut f{\"u}r Computing NIC, Deutsches Elektronen-Synchrotron DESY,\\
Platanenallee 6, 15738 Zeuthen, Germany}
\affiliation[h]{Institut f{\"u}r Physik, Humboldt-Universit{\"a}t zu Berlin\\
Newtonstr. 15, 12489 Berlin, Germany}

\affiliation[ma]{Instituto de F\'{\i}sica Te\'orica UAM-CSIC and Dpto. de F\'{\i}sica Te\'orica\\
C/~Nicol\'as Cabrera 13-15, Universidad Aut\'onoma de Madrid, Cantoblanco E-28049 Madrid, Spain}

\affiliation[mu]{Universit\"at M\"unster, Institut f\"ur Theoretische Physik,\\
Wilhelm-Klemm-Stra{\ss}e 9, 48149 M\"unster, Germany}

\affiliation[mar]{Aix-Marseille Universit\'e, Universit\'e de Toulon, CNRS, CPT, Marseille, France
}

\affiliation[d]{School of Mathematics, Trinity College Dublin, Dublin 2, Ireland}

\affiliation[c]{Theoretical Physics Department, CERN, 1211 Geneva 23, Switzerland}

\affiliation[mz]{Helmholtz Institute Mainz, Johannes Gutenberg University, Mainz, Germany}
\affiliation[g]{GSI Helmholtz Centre for Heavy Ion Research, Darmstadt, Germany}

\emailAdd{rainer.sommer@desy.de}

\abstract{In a somewhat forgotten paper \cite{Guazzini:2007ja} it was shown how to perform interpolations between relativistic 
and static computations in order to obtain  results for heavy-light observables for masses from, say, $m_{\rm charm}$ to $m_{\rm bottom}$.
All quantities are first continuum extrapolated and then interpolated in $1/m_h=1/m_{\rm heavy}$. Large volume computations are combined with
finite volume ones where a relativistic bottom quark is accessible with small $am_{\rm bottom}$. 
We discuss how this strategy is extended to semi-leptonic form factors and other quantities of phenomenological
interest. The essential point is to form quantities where the limit $m_h\to\infty$ is approached with power corrections $\rmO(1/m_h)$ only. Perturbative corrections $\sim\alpha_s(m_h)^{\gamma+n}$ are cancelled in the construction of the observables. We also point out how such an approach can help to control systematics in semi-leptonic decays with just large volume data. 
First numerical results with $N_f = 2 + 1$ and lattice spacings down to 0.039 fm are presented in \cite{alessandro}.
}

\FullConference{%
The 40th International Symposium on Lattice Field Theory,\\
31st July-4th August, 2023,\\
Fermilab, Batavia, Illinois, USA
}

\usepackage{defs}

\begin{document}
\maketitle

%%%%%%%%%%%%%%%%%%%%%%%%%%%%%%%%%%%%%%%%%%%%%%%%%%%%%%%%%%%%%%%%%%%%%%%%%
\section{Introduction}
\label{s:intro}
B-physics is a possible portal to physics beyond the Standard Model. This is even 
more relevant now than in the past since direct searches have not provided evidence for new physics.  Small tensions between standard model and experiment exist in B-physics, but others may be hidden by large uncertainties or an inaccurate treatment of the theory. Indeed, precise theory predictions, e.g.\ for B-meson decays are very challenging, because perturbation theory seems, generically, inaccurate (see below) and non-perturbative 
computations on the lattice are complicated by the large ratio of scales 
between the b-quark mass and the non-perturbative scales of QCD 
including the pion mass. 

\begin{figure}
\centering
	\includegraphics[width=0.8\textwidth]{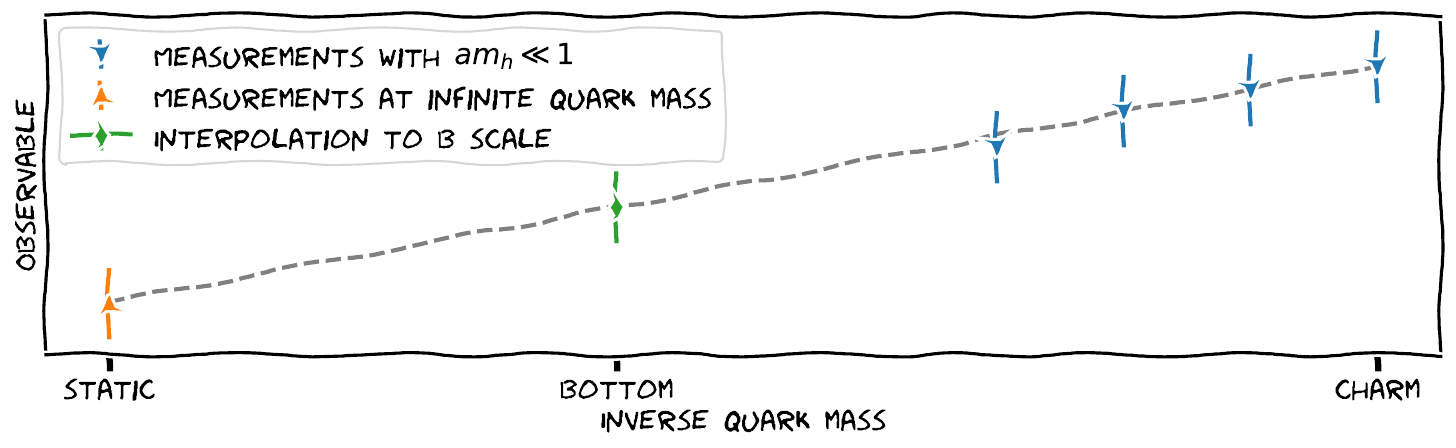}
	\caption{Sketch of interpolation between static and and relativistic data in the continuum.
	\label{f:sketch}
}
\end{figure}

Because of the large scale ratio, all existing computations for B-physics on the lattice make use of expansions in $1/m_b$, i.e.\  effective field theories (EFTs),
in one way or the other \cite{FlavourLatticeAveragingGroupFLAG:2021npn}. Also so-called relativistic computations constrain the quark mass dependence of heavy-light observables to forms motivated by an expansion in $1/m_b$ \cite{El-Khadra:1996wdx,Aoki:2001ra,Christ:2006us,ETM:2016nbo,Bazavov:2017lyh,Boyle:2018knm,Hatton:2021syc,Colquhoun:2022atw,Flynn:2023nhi}. It is important to reduce the number and importance of assumptions made in the EFT treatment to a minimum. A  step was taken a while ago~\cite{Guazzini:2007ja}, combining 
the relativistic theory,
\begin{equation}
	\lag{} =\lag{\mathrm{glue}}+\lag{\mathrm{light}} +\lag{\mathrm{h}}\,,\quad\lag{\mathrm{h}} = \bar \psi_{h} ({D_\mu}\gamma_\mu + m_{h})\psi_{h} \,,
\end{equation}
with the static effective theory \cite{Eichten:1989zv} ($\heavyb\frac{1+\gamma_0}2=\heavyb\,,\quad   
	\frac{1+\gamma_0}2\heavy=\heavy$),
\begin{equation}
\label{e:Lstat}
	\lag{\mathrm{eff}} =\lag{\mathrm{glue}}+\lag{\mathrm{light}} +\lag{\mathrm{stat}}\,,\quad\lag{\mathrm{stat}} = \heavyb (D_0 + \mhbare)\heavy\,.
\end{equation}
 %($\lag{b} = \heavy D_0 \heavy $)
The continuum limit is taken separately in both theories, 
which in practice, due to accessible $a$, requires $m_h \lessapprox m_b/2$ in the relativistic case. Results for the physical quark mass, $m_b$, are obtained by interpolation in $1/m_h$, where the static effective theory yields a point at $1/m_h=0$, see \cref{f:sketch}.
 
The above work~\cite{Guazzini:2007ja} built on the introduction of step scaling in the static approximation \cite{Heitger:2001ch} and in HQET \cite{Heitger:2003nj} as well as for finite mass heavy quarks \cite{Guagnelli:2002jd,deDivitiis:2003iy}. However, the applications to semi-leptonic decays were complicated \cite{deDivitiis:2007ptj}. Here we discuss that  semi-leptonic decays and more can be included by a very simple generalisation. Instead of focusing on the
$1/m_h$ expansion of finite volume effects and step scaling as done in the past, the general strategy rests solely on the basic requirement for a combination of static and relativistic results: observables have to admit a simple interpolation in $1/m_h$ as in \cref{f:sketch}.

%%%%%%%%%%%%%%%%%%%%%%%%%%%%%%%%%%%%%%%%%%%%%%%%%%%%%%%%%%%%%%%%%%%%%%%%%
\section{General strategy}
\label{s:general}

Our general strategy is then to form {\bf suitable quantities} which 
\renewcommand{\labelenumi}{\roman{enumi}.}
\begin{enumerate}[parsep=-4pt]
	\item can be computed in the continuum limit,
	\item can be combined to obtain the desired observables, e.g.\ decay constants or semi-leptonic form factors,
	\item possess a simple behavior as a function of $1/m_h$, such that they can be interpolated in this variable.
\end{enumerate}
We need low energy quantities such that the $1/m_h$ expansion is applicable. Furthermore, the static limit $m_h \to \infty$ has to exist with no logarithmic corrections but only power corrections $\rmO(1/m_h)$. In our context, logarithmic corrections means $\sim \alpha^{n+\gamma}(m_h)$ while power corrections may (and will) contain logs: $\alpha^n(m_h)/m_h=\rmO(1/m_h)$.

The step scaling strategy of \cite{Guazzini:2007ja} satisfies these requirements, but it is  too restrictive. As a preparation for  examples how  a general strategy can be applied, we discuss the all-important renormalisation in the static theory. 

\subsection{Renormalisation and matching in the static theory}
\label{s:LV}
While it is not proven, %to all orders in the coupling expansion, 
the static effective theory is renormalisable according to all that we know. For a discussion and references we refer to sect.~1.6 of \cite{Sommer:2015hea}. Beyond renormalisation any effective theory has to also be matched to the fundamental one, such that the observables agree up to power corrections, here $\rmO(1/m_h)$. These two steps are cleanly separated by using renormalisation group invariants. We briefly recall the facts which we need later, see e.g.\ \cite{Sommer:2015hea} for details.

Energies defined and computed with the static Lagrangian \cref{e:Lstat} are finite after additively renormalising its bare mass,
\begin{equation}
\label{e:mbare}
	\mhbare = \mathrm{finite} + \delta m\,, 
	\quad  \delta m = \frac1a [c_0 g_0^2 + \ldots]\,.
\end{equation} 
Since  all energies of states with the quantum number of a single b-quark have the property
\begin{equation}
	E^\mathrm{stat}_n(\mhbare) = E^\mathrm{stat}_n(0)+\mhbare\,,
\end{equation}
energy differences $E_n^\mathrm{stat}-E_m^\mathrm{stat}$ are finite and independent of the finite piece in \cref{e:mbare}. 
These differences then satisfy our criteria i.-iii. provided the energies correspond to low lying states and momenta are small.

Furthermore there is considerable interest in transition matrix elements of local electroweak operators, e.g.\ $L_\mu=V_\mu-A_\mu$, %,\; \op{\Delta b =2}^\mathrm{L}$, 
where $V_\mu,A_\mu$ are heavy-light currents.  Depending on whether parity changes or not, only $A_\mu$ or $V_\mu$ contribute.  
These local fields are renormalised and matched  multiplicatively,
\begin{equation} \label{e:multiplicative}
	\op{\mathrm{stat}}(x;m_h^\mathrm{RGI}) = C_{\op{}}(m_h^\mathrm{RGI}/\Lambda)\, \op{\rm stat}^\mathrm{RGI}(x) \,,
	\quad \,\op{\rm stat}^\mathrm{RGI}(x)=Z_{\op{}}(g_0)\, \op{\rm stat}^\mathrm{bare}(x;g_0)\,.
\end{equation}
Note that in the static limit Lorentz symmetry is of course broken and, e.g.\, $\op{}=V_0$ and  $\op{}=V_k$ have different factors $C_{V_0} \ne C_{V_k}$, see \cite{Sommer:2010ic}. Depending on the lattice regularisation one may also have $Z_{V_0} \ne Z_{V_k}$. An important fact is  that
\begin{equation}
	C_{\op{}}(m_h^\mathrm{RGI}/\Lambda) \simas{m_h\to \infty} \;[\log(m_h^\mathrm{RGI}/\Lambda)]^{\hat \gamma_{\op{}}}
\end{equation} 
is not finite in the static limit.\footnote{Also the renormalisation factors
\begin{equation}
	Z_{\op{}}(g_0) =1 + Z_1^{\op{}}g_0^2 + Z_2^{\op{}}g_0^4 +
	\ldots
\end{equation}
have a logarithmic dependence on $a$ since
$g_0^2 \sim -1/\log(a\Lambda)$, but it is sub-leading. }
 Therefore 
matrix elements of $\op{\mathrm{stat}}$ by themselves are not ``suitable quantities''.
Instead, ratios of different matrix elements of the same 
operator and with the same $m_h$ can be used.
In the old step scaling method \cite{Guazzini:2007ja,deDivitiis:2007ptj,Guagnelli:2002jd},
the space-time volume and only it changed between the two matrix elements that form the ratio. This leads to a rather impractical scaling of the momenta for form factors. That restriction is unnecessary. 

We turn to examples implementing the general strategy.

\subsection{Large volume}
\label{s:LV}
The form factors for $B\to \pi \ell \nu$ decays are very relevant for the determination of $V_\mathrm{ub}$, the test of CKM unitarity as well as the search for deviations from the Standard Model. The  QCD matrix elements of the vector current determine two independent form factors, denoted by $h_\parallel$ and $\;h_\perp$ in the HQET basis.
Labelling the hadronic states by the spatial momenta of the hadrons and using non-relativistic normalisation\footnote{The normalisation  conditions in a finite spatial $L\times L\times L$ volume are 
\begin{eqnarray}
\langle B(\vec{p})|B(\vec{p}^{\,\prime})\rangle \,&=&\, 2L^3 \delta_{\vec{p},\vec{p}^{\,\prime}}\,,
\quad
\langle B^*(\vec{p},\lambda)|B^*(\vec{p}^{\,\prime},\lambda^{\,\prime} )\rangle \,=\, 2L^3 \delta_{\vec{p},\vec{p}^{\,\prime}}\,\delta_{\lambda,\lambda'},
\\
\langle \pi(\vec{p})|\pi(\vec{p}^{\,\prime})\rangle \,&=&\, 2E_{\pi,\vec{p}}L^3 \delta_{\vec{p},\vec{p}^{\,\prime}}\,,
\end{eqnarray}
with $\lambda$ the polarisation of the vector meson state.
} for the B-meson state, the form factors are defined by
($E^2_\pi=\vec p_\pi^2+m_\pi^2$)
\begin{equation}
	\langle\pi(\vec p_\pi)| {V}^k(0)
	| B(\vec{0})\rangle 
	=\sqrt{2}p^k_\pi \,h_\perp(E_\pi)\,, \quad 
	\langle\pi(\vec p_\pi) | {V}^0(0)
	|B(\vec{0})\rangle =\sqrt{2}\,h_\parallel(E_\pi)\,.
\label{e:formfacts}
\end{equation}
Normalising by the form factor at a reference energy $E_\pi^\mathrm{ref}$, we define %the observable
\begin{equation}
  \tau_x = \log\left(h_x(E_\pi) / h_x(E_\pi^\mathrm{ref})\right) 	\,,\quad x\in\{\perp,\parallel\}\,,
\end{equation}
where in the static theory matching and renormalisation factors cancel,
\begin{equation}
	\lim_{m_h\to\infty} \tau_x = \tau_x^\mathrm{stat} = 
	\log\left(h_x^\mathrm{stat}(E_\pi) / h_x^\mathrm{stat}(E_\pi^\mathrm{ref})\right) =
	\log\left(h_x^\mathrm{stat,bare}(E_\pi) / h_x^\mathrm{stat,bare}(E_\pi^\mathrm{ref})\right).
\end{equation}
More precisely, the static limit is approached with power corrections only,
\begin{equation}
    \tau_x = \tau_x^\mathrm{stat} 
	+ \rmO(E_\pi/m_h, \Lambda/m_h)\,,
	\label{e:ffratio}
\end{equation}
and i) as well as iii) are satisfied.
The strategy now consists in the following steps.
\renewcommand{\labelenumi}{\alph{enumi})}
\begin{enumerate}[parsep=-4pt]
\item \label{it:a}	Compute the lhs of \cref{e:ffratio} in the relativistic theory for $am_h\ll 1$, such that a standard Symanzik analysis applies and take the continuum limit. In practice one is limited to  $m_h\lessapprox m_b/2$ or similar.
\item A separate continuum limit in the static theory yields $\tau_x^\mathrm{stat}$. 
\item \label{it:c}
      The b-quark mass scale can be reached by an interpolating fit, e.g.\ 
\begin{equation} 
\label{e:fitfct}
	\tau_x^\mathrm{fit} = \sum_{n=0}^{N}t_{x,n} \left(\frac{E_\pi}{\mps}\right)^n
\end{equation}  
to all continuum results, including $\tau_x^\mathrm{stat}$. Here  the heavy-light pseudo-scalar  mass, $\mps$, is used as a proxy for the heavy quark mass and the fit function is finally evaluated at $\mps=m_B$ where $m_h$ 
becomes $m_\mathrm{b}$.
\end{enumerate}
Eq.~(\ref{e:fitfct}) is a model since logarithmic modifications of the power corrections 
 ($n\geq1$) are neglected and one has to truncate at some order $N$. However, since we do an interpolation and not an extrapolation the model dependence  
  is expected to be small, but this will have to be assessed case-by-case. The leading $n=0$ term is free of logs by construction.
 
 We end this part by a modification which we expect to be useful in practice. It is always good to have normalising factors defined in such a way that they are numerically very precise. Here this suggests $\vec p^\mathrm{ref}=\vec 0$ or equivalently $E_\pi^\mathrm{ref}=m_\pi$ We recommend this for $\tau_\parallel$, but the form factor $h_\perp$
 is not defined for vanishing momentum. Hence, we propose to switch to a different matrix element of the spatial vector current. The vector meson decay constant,
 \begin{equation}
 	\hat f_V \epsilon_k^\lambda =  \langle{B^*(\vec0,\lambda)} |{V}^k(0)|0\rangle\,,
 \end{equation}
 can be determined precisely  from  zero
 momentum two-point functions of the vector current. We thus propose to use
 \begin{equation}
  \tau_\parallel = \log\left(h_\parallel(E_\pi) / h_\parallel(m_\pi)\right) 	\,,\quad \tau_\perp = \log\left(E_\pi h_\perp(E_\pi) /(\Lref \hat f_V)\right)\,
\end{equation}
in practice. The factors $E_\pi$ and $\Lref$ get the mass dimensions straight and, in this particular case, the  choice $\Lref=1/f_\pi$ 
is motivated by lowest order HMChPT where $
E_\pi f_\pi h_\perp(E_\pi) /\hat f_V= g_{BB^*\pi} /\sqrt{2}$ \cite{Burdman:1993es}.
   
\subsection{Step scaling}
\label{s:SS}

With the above, the $E_\pi$-dependence ($q^2$ in a general frame) of the form factors can be computed and compared to experiments. However, the absolute normalisation is lost. A determination of $|V_{ub}|$ does also need  the absolute normalisation. We thus also want a strategy for the computation of $\hat f_V$ and $h_\parallel(m_\pi)$. It is provided by the step scaling approach of \cite{Guazzini:2007ja}. 
Compared to the strategy above, the energy variable is replaced by the extent of a finite volume and a crucial point is that one can compute quantities directly in the relativistic theory when the size of the volume is of the order of $L_1\approx 0.5\fm$, where lattice spacings are accessible such that  $a m_b \ll 1$. The matching factors $C_x(m_b^\mathrm{RGI}/\Lambda)$ are then replaced by a direct relativistic computation and step scaling functions independent of $C_x$ transport this information to large volume. 
A slight complication is that in finite volume we need a non-perturbative proxy for the quark masses. The natural choice is a finite volume heavy-light ``mass'' $\mps(L)$ with the property $\lim_{L\to\infty} \mps(L)=\mps$. The first step therefore is the 
determination of the finite-size dependence of $\mps(L)$. 

We introduce one length-scale $\Lref$ to form dimensionless observables and write (in principle more steps $\sigma$ can be inserted) 
\begin{eqnarray}
	\Lref \mB &=& \Lref \mps(L_1) + \sigma_m(u_1,y_2) + \rho_m(u_2,y_B)\,,
\label{e:mHss}
	\\
	\sigma_m(u_1,y_2) &=&\Lref[\mps(L_2)-\mps(L_1)]\,,\quad \rho_m(u_2,y_B) = \Lref[\mB-\mps(L_2)]\,.
\end{eqnarray}
Here all quantities refer to the same heavy quark mass
and the light quark masses are set to zero in finite volume. Since the functions $\sigma_m$ and $\rho_m$ are differences of energies, they are finite in the static theory; $\delta m$ drops out. Steps a)-c) can therefore be carried out for $\rho,\sigma$ as written down for $\tau$. The variables
$u_i$ are proxys for the sizes $L_i$ in the form of values of running couplings, $u_i=\bar g^2_\mathrm{GF}(L_i)$ \cite{DallaBrida:2016kgh} and
the variables $y_i=\Lref\,\mps(L_i)$ are proxys for the b-quark mass.
They are obtained recursively going from large volume to small,  
\begin{equation}
	y_\mathrm{B}\equiv\Lref\,\mB\,,
	\quad y_2=y_\mathrm{B}-\rho_m(u_2,y_B)\,,
	\quad y_1=y_2-\sigma_m(u_1,y_2)\,.
\end{equation}
Following this chain imposes that the quark mass is set to the physical b-quark mass.
Of course, starting from a different input, e.g.\ $m_B\to m_D$, one uses a different heavy quark mass in all subsequent steps.

\subsubsection{Quark mass}
When small lattice spacings are available and a non-perturbative renormalisation of the quark mass is carried out~\cite{Campos:2018ahf}, the function 
\begin{equation}
	\pi_m(u_1,y) = \frac{\mps(L_1)}{m_h^\mathrm{RGI}} 
	=\frac{y}{\Lref m_h^\mathrm{RGI}}
\end{equation}  
can be computed in the continuum limit. Combined with
$y_1$ it yields the renormalisation group invariant b-quark mass
\begin{equation}
	m^\mathrm{RGI}_\mathrm{b} = \frac1{\Lref}\frac{y_1}{\pi_m(u_1,y_1)}\,. 
\end{equation}

\subsubsection{Decay constants and other multiplicatively renormalised matrix elements}
As discussed above, the vector meson decay constant 
is likely to play an important role in a precise determination of  $b\to u $ semi-leptonic decays
and the pseudo-scalar one  yields a relevant
crosscheck through leptonic $B$ decays.

Our strategy for their determination parallels the one for the quark mass and uses the already known quark mass proxys $y_i$. Decay constants are matrix elements of the currents $V_\mu$ and $A_\mu$ which
are multiplicatively renormalised and matched in the static theory. We obtain a basic equation which is form-identical to (\ref{e:mHss}) by defining the finite and infinite volume observables to be the logarithms of the dimensionless matrix elements. We write it generically for any multiplicatively renormalised matrix element, made dimensionless by our reference scale $\Lref$,
\footnote{\label{foot}We here focus on  matrix elements of a single operator. There are also cases where a single operator in QCD is mapped to more operators in the static theory which mix under renormalisation and under matching. Furthermore, mixing can already be present in the relativistic theory. Then \cref{e:multiplicative} holds with a $\Nop$-component vector $\op{}$  and $\Nop\times \Nop$ matrices $C,Z$. We then need to consider $\Nop$ different matrix elements of the operators; $\me$ is $\Nop\times \Nop$ with the first index corresponding to the different operators  and the second one to the different matrix elements. The master equations (\ref{e:mechain}-\ref{e:sigmarho}) are changed to
\begin{eqnarray}
	\me_\infty &=&  \rho_\me(u_2,y_B) \;\sigma_\me(u_1,y_2) \;\me(L_1) \,,
	\\
	&&\quad\rho_\me(u_2,y_B) = \me_\infty\,\me(L_2)^{-1} \,,\quad \sigma_\me(u_1,y_2) = \me(L_2)\,\me(L_1)^{-1}\,.
\end{eqnarray}
}
\begin{eqnarray}
	\log\left(\me_\infty\right)  &=& \log\left(\me(L_1)\right) + \sigma_\me(u_1,y_2) + \rho_\me(u_2,y_B)\,,	
	\label{e:mechain} 
	\\
	&&\sigma_\me(u_1,y_2) = \log\left(\me(L_2)\,/\,\me(L_1)\right)\,,\; \rho_\me(u_2,y_B) = \log\left(\me_\infty\,/\,\me(L_2)\right)\,.
	\label{e:sigmarho}
\end{eqnarray}
In particular, for the {\bf vector decay constant},
we choose in our {\bf practical implementation} %of the strategy 
\cite{alessandro} 
\begin{eqnarray}
&& ~~\includegraphics[width=3cm,trim = 0cm 13mm 0mm 0mm]{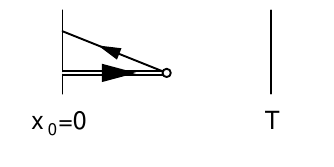}~~
\nonumber
\\
	\me(L)\propto\hat f_V(L) \propto \langle{B^*(\vec0,\lambda);L} |{V}^k(0)| \Omega;L\rangle 
	&=& \text{-------------------------------}\,\;.
	\\
&& \left(\includegraphics[width=3cm,trim = 0cm 14mm 0mm 0mm]{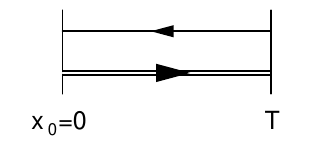}\right)^{1/2} 
\nonumber
\end{eqnarray}
Here the states are finite volume states, $| \Omega;L\rangle$ has  vacuum quantum numbers, $|B^*(\vec0,\lambda);L \rangle$ is a zero momentum  $B^*$-meson state with polarisation $\lambda$. The matrix element is constructed from  correlation functions with Schr\"odinger functional boundary conditions \cite{Luscher:1992an,Sint:1993un} as indicated in the equation.
A precise definition is $\log(L^{3/2}\hat f_V(L)/\sqrt{2})=\Phi_{19}$ with $\Phi_{19}$ of  \cite{DellaMorte:2013ega}, but with the projection to zero topological charge as for the coupling~\cite{DallaBrida:2016kgh}.

%%%%%%%%%%%%%%%%%%%%%%%%%%%%%%%%%%%%%%%%%%%%%%%%%%%%%%%%%%%%%%%%%%%%%%%%%
\section{Lines of Constant Physics and continuum limits}
Lines of Constant Physics (LCP) specify how bare parameters are scaled as the resolutions $a/L_i$ of the lattice theory change. 
They can  differ from quantity to quantity. Indeed, it is useful to define the LCP adapted to the physics 
involved. 

The overall scale $\Lref$ could be taken e.g.\ as the 
gradient flow (GF) scales $\sqrt{t_0}$ \cite{Luscher:2010iy} or $w_0$ \cite{BMW:2012hcm}. Since we work with finite volumes, it is more natural to define the scale through the GF running coupling in
one of the used volumes. In our numerical example \cite{alessandro} practical considerations lead to the choice \cite{Fritzsch:2018yag}
\begin{equation}
	\Lref = 4L_0\,, \quad \bar g^2(L_0) = u_0=3.949\,,\quad [L_1=2L_0,\; L_2=4L_0=\Lref] \,,
\end{equation}
in terms of the GF coupling defined exactly as in \cite{DallaBrida:2016kgh}.
The arguments $u_i=\bar g^2(L_i)$ of our  scaling functions are
then given by the precisely known \cite{DallaBrida:2016kgh} coupling step scaling functions
\begin{equation}
	u_1=\sigma(u_0),\,\quad u_2=\sigma(u_1)\,.
\end{equation}
Taking the knowledge of $\sigma(u_i)$ for granted,  we need three LCP's for the above strategy. 
In principle, they only differ by the 
values of the renormalised coupling $u$ and the proxy for the heavy quark mass $y$. For the latter one chooses a set of values compatible with $am_h\ll1$.
The LCPs are the set of conditions
\begin{equation}
	\lcp_0= \left\{\bar g^2(g_0,  L/a)=u, \quad 
	\frac{L}{a} \Mps(g_0, am_h, L/a)=y \,,\quad  m^\mathrm{PCAC}_l=0, \; l=1\ldots\nf\,\right\}\,,\quad \Mps=a\mps\,,
\end{equation}
where the dimensionless arguments are the bare coupling, $g_0$, bare %(quenched) 
heavy quark mass, $am_h=1/(2\kappa_h)-4$, and resolution $L/a$. The bare light quark masses are  fixed by the last $\nf$ conditions. 
The lattice approximants for the step scaling functions
(we use the special case $\Lref=2 L_1$) are
\begin{eqnarray}
		\Sigma_m(u,y,a/L) &=& 2L/a \,\left[\Mps( g_0, am_h, 2L/a) - \Mps( g_0, am_h, L/a)\right]_{\lcp_0}\,,\quad \, 	
		\\
	\Sigma_{\me}(u,y,a/L) &=& 
	\log\left(\me( g_0, am_h, 2L/a)\,\me( g_0, am_h, L/a)^{-1}\right)_{\lcp_0}
	\,,
\end{eqnarray}
where we are interested in $u=u_1$ and a range of $y$. 

Analogously, the small volume function is given by 
\begin{equation}
	\pi_m(u,y) = \lim_{a\L\to0} \Pi_m(u,y,a/L)\,,\quad \Pi_m(u,y,a/L)=
	\left. \frac{\Mps( g_0, am_h, L/a)}{M^\mathrm{RGI}_h( g_0, am_h, L/a)}
	\right|_{\lcp_0}
	 \,,\quad M^\mathrm{RGI}_h = am^\mathrm{RGI}_h \,.
\end{equation}  
In large volume, massless light quarks are not desired. 
We therefore introduce a massive LCP, $\lcp_\mathrm{m}$. It may correspond to the physical point of isoQCD~\cite{FlavourLatticeAveragingGroupFLAG:2021npn} or the symmetric point of CLS ($m_\pi=m_\mathrm{K}\approx410\,\mathrm{MeV})$ \cite{Bruno:2017gxd}). It is implied that the volume is large enough such that finite size effects are negligible. $\lcp_\mathrm{m}$ applies straight forwardly to the pure large volume quantities $ \me = \lim_{a/\Lref \to0} \,\me(g_0,am_h,am_i)$, while for the functions connecting large volume and finite $L$ one has, e.g.\ 
\begin{equation}
	\rho_\me(u,y) = \lim_{a\L\to0} \Rho_\me(u,y,a/L)\,,\quad \Rho_\me(u,y,a/L)=
	\log\left( \frac{
	\me( \tilde g_0, a\tilde m_h, am_i)_{\lcp_\mathrm{m}}}{
	\me( \tilde g_0, a\tilde m_h, L/a)_{\lcp_0}}
	\right)
	  \,,	
\end{equation}
where $\lcp_0$ defines the parameters in the denominator and $\lcp_\mathrm{m}$ in the numerator. The improved bare parameters
$\tilde g_0, a\tilde m_i$ are defined in~\cite{Luscher:1996sc,Bhattacharya:2005rb}. Since the light quark masses vanish for $\lcp_0$, we have $\me( \tilde g_0, a\tilde m_h, L/a)_{\lcp_0} = \me( g_0, am_h, L/a)_{\lcp_0}$.

\section{Conclusion and Outlook}

We have pointed out that the core idea of \cite{Guazzini:2007ja} is applicable beyond the step scaling in 
volume. Extrapolations of the relativistic theory to the b-quark mass can be turned into interpolations in any situation where the static  theory result is free from logarithms 
in the heavy quark mass $m_h$. Generically such logs are present. They originate from loop-corrections in the matching of the effective theory to QCD. Because of the simple structure of the static theory it is easy to find functions where all $m_h$-dependence cancels. For matrix elements of local operators one takes ratios of matrix elements between different states.
This means that the simple interpolation of \cref{f:sketch} can determine the energy dependence of the form factors, but not the normalisation. The latter becomes accessible by adding step scaling in the volume. A single step with just one $\sigma$ appears to be enough and first results are encouraging~\cite{alessandro}.  

The somewhat involved step scaling part has to be done only once but including a continuum limit. The result can then be used with any action and one can concentrate on the particular challenges appearing in large volume, e.g.\ controlling excited state effects \cite{Bar:2023sef} and form-factor parameterisations \cite{Flynn:2023qmi,Di_Carlo_2021}.

We note that the strategy is applicable beyond the explicit cases discussed. E.g. \cref{foot} applies to the $2\times2$ mixing problem encountered in $B\bar{B}$ mixing in the standard model with twisted mass fermions or with exact lattice chiral symmetry~\cite{Palombi:2006pu}.

\vspace{0.5em}
\begin{small}
\noindent
We thank Oliver B\"ar, Alexander Broll and Andreas J\"uttner for discussions and AJ also for comments on a first version of the text.
We acknowledge support from the EU projects EuroPLEx
H2020-MSCAITN-2018-813942 (under grant agreement No.~813942),
STRONG-2020 (No.~824093) and HiCoLat (No.~101106243),
as well as from grants,
PGC2018-094857-B-I00, PID2021-127526NB-I00, SEV-2016-0597,
CEX2020-001007-S funded by MCIN/AEI,
the Excellence Initiative of Aix-Marseille University - A*Midex
%a French “Investissements d’Avenir” programme,
(AMX-18-ACE-005),
and
 from the Deutsche Forschungsgemeinschaft, (GRK 2149 and GRK2575).
J.H. thanks the Yukawa Institute for Theoretical Physics, Kyoto University, for its hospitality.
\end{small}

\bibliographystyle{JHEP}
\bibliography{proceedings.bib}

\end{document}